\documentclass[12pt,preprint]{aastex}

\begin{document}
\title{Anthropic Argument for Three Generations}

\author{Andrew Gould}
\affil{Department of Astronomy, Ohio State University,
140 W.\ 18th Ave., Columbus, OH 43210, USA; 
gould@astronomy.ohio-state.edu}

\begin{abstract}
The standard model of particle physics contains $N_{\rm gen}=3$ generations
of quarks and leptons, i.e., two sets of three particles in each
sector, with the two sets differing by 1 unit of charge in each.
All 12 ``predicted'' particles are now experimentally accounted for, and 
there are strong (though not air-tight) arguments that there are no 
more than three generations.  The question is: why exactly $N_{\rm gen}=3$?
I argue that three generations is a natural prediction of the
multiverse theory, provided one adds the additional, quite reasonable
assumption that $N_{\rm gen}$ in a randomly realized
universe is a steeply falling function of number.  In this case
$N_{\rm gen}>2$ to permit CP violation (and so baryogenesis 
and thus physicists) and $N_{\rm gen}<4$
to avoid highly improbable outcomes.
I thereby make a testable anthropic-principle prediction:
that when a theory of randomly realized $N_{\rm gen}$ is developed,
the probability will turn out to be steeply falling in $N_{\rm gen}$.

\end{abstract}


\section{{Introduction}
\label{sec:intro}}

After Anderson and Neddermeyer's 1936 discovery of the muon was confirmed
by \citet{street37}, I.~I.~Rabi famously quipped ``who ordered that?'',
i.e., why was there a second ``electron''?  No sensible answer to this
question could even be attempted until the general pattern of
``multiplicity'' of ``fundamental'' particles was established.

The emergence of a standard particle physics model does allow this
question to be at least properly framed.  In this model, there
are exactly three ``electrons'' (electron, muon, tau), 
and each is associated with its corresponding neutrino, with 
identical quantum numbers except 1 extra unit of charge.  In 
parallel, there are exactly three ``lower quarks'' (down, strange, bottom),
each with its corresponding ``upper quark'' (up, charm, top), also
with identical quantum numbers except 1 extra unit of charge.
The standard model has demonstrated at least some predictive
power (as opposed to being merely a post-facto classification
scheme) because the top quark was firmly established in the model well
before its experimental confirmation.

With four classes of particles in each of three ``generations'', there
are 12 predicted particles.  
All 12 members of these three generations have been confirmed experimentally
(so $N_{\rm gen}\geq 3$), 
and there is a powerful piece of evidence that there are no more than three 
generations.  The $Z$ particle can decay into particle/anti-particle
pairs of any of these 12 particles, except for the top quark, since
two tops have more mass than  the $Z$.
The rate of decay would increase (and so the width of
the $Z$ resonance would decrease) beyond its measured value if there were
particles in a fourth generation.  The only caveat is that if all four
particles in this putative generation were heavier than half the $Z$
mass, these decay channels would be blocked (as they are for the top
quark).

Hence, there is excellent, though not absolutely secure evidence
that there are exactly three generations ($N_{\rm gen}=3$).
And so, Rabi's question can
now be made more precise: ``why exactly three generations''?

\section{{In Defense of the Anthropic Principle}
\label{sec:anthropic}}

There is a broad class of answers to such questions that is
subsumed under the lofty slogan ``anthropic principle''.
The core idea of this principle is that our ``universe'' is only
one of many universes, each with its own ``fundamental constants'',
such as the electron mass, the fine structure constant, etc.
These constants appear as ``fundamental'' (i.e., without any
further explanation -- or perhaps ``explained'' by mathematical
derivation from other constants that are themselves unexplained),
but they actually are just realizations of fields whose symmetries
are broken as the universe cools, leaving them at some
random value.  Then there are a huge number of universes that
have various values for these constants that are incompatible with
intelligent life, and so do not contain physicists to ponder the
values of these constants.  Our universe is among the others.
Hence, if we see that certain constants (or combinations of constants)
``happen'' to be compatible with life, the reason is the same as why
the Earth ``happens'' to have water: our planet may well be in a minority
that are so endowed, but the others do not have people on them to
worry about this issue.

Of course, the full conditions for intelligent life are not known,
but we can conservatively identify at least some conditions.  For
example, if big bang nucleosynthesis had ended with $>99\%$ helium,
then stars would not live long enough for intelligent life to evolve,
even supposing that such life could form without hydrogen.  And I think
that few would argue that a universe without baryons 
(protons and neutrons -- made of quarks) could contain
life, intelligent or otherwise.

Now, before continuing, I must take note of the fact that many
people object to the ``anthropic principle'' on the grounds that
it is ``not a scientific theory'' in that it ``does not make
testable predictions''.  Such arguments reflect a deep
misunderstanding of the nature of scientific inquiry. 
Of course, the anthropic principle is
not a scientific theory and obviously in itself makes no testable predictions.
Rather it is a framework for theoretical speculation.  Any profoundly
new theory will be preceded by theoretical speculation, or even
groping, before it can be properly formulated.  Such full formulations
may require additional universes and at the same time make predictions
about our universe.  If there are one or two such predictions that
are verified, and these are minimally entangled with the hypothesis
of other universes, one might maintain the hope that a new theory will
emerge that predicts the same things about our universe but avoids
the ``embarrassment'' of other universes.  But if these correct predictions
multiply, and if they become deeply entangled with the existence
other universes, then the other universes will come to be accepted,
in the same way that we currently accept the ``reality'' of the magnetic
vector potential, despite the fact that it was originally introduced
as a mathematical convenience.  Of course, it is also possible that
nothing will come of the anthropic-principle speculation, in which
case it would join the ranks of the vast majority of such speculations
in the waste bin of theoretical physics.

In this context, it is useful to catalog physical constants that
would be (post-facto) explained by the anthropic principle, assuming
that many universes with very different physical constants do exist.

\section{{The Anthropic Explanation of Three Generations}
\label{sec:explanation}}

In 1967, the great Soviet physicist Andrei
Sakharov identified three conditions for baryogenesis.  In the
early universe, there were exactly equal numbers of baryons and
anti-baryons, and these both approximately equaled the number of
photons.  Today, the number of photons is roughly unchanged, but
essentially all of the anti-baryons have annihilated with baryons.
 From the presently observed baryon/photon ratio, we therefore learn
that somehow during those early times, about one
in a billion anti-baryons was converted into a baryon.
Sakharov's (1967) three necessary conditions were 1) baryon-number
violating process,
2) violation of charge-parity (CP) symmetry, 3) out-of-equilibrium
thermodynamics.

The first condition is obvious.  The third is also obvious, since
in thermodynamic equilibrium detailed balance ensures that every
baryon-violating process will be countered by baryon violation
going in the other direction.  The second is less obvious.
Under the CP symmetry, a given particle's anti-particle will behave
exactly as the particle does, provided we consider anti-particles
of the opposite parity.  In quantum mechanics, CPT symmetry is
essentially a mathematical identity.  That is, the above symmetry
must hold if, in addition, the anti-particle is going backwards through time.
Hence, breaking the CP symmetry is essential to breaking symmetry
in time, which is required to move from a state of 0 baryon number
to positive baryon number at a later time.

Sakharov was inspired to consider this problem by Cronin and Fitch's discovery
of CP violation in the neutral kaon system \citep{christenson64}.
Neutral kaons are composed of quarks from the first two generations,
down and strange.  But, if there were only two generations, CP
violation would be mathematically impossible: the matrix linking
the mass states and the flavor states of these particles could
always be ``rotated'' so that the CP violating terms were zero.
Realizing this, \citet{kobayashi73} introduced a third generation
of quarks (and the Cabbibo-Kobayashi-Maskawa matrix to link them)
to explain CP violation, even though no such third generation
had yet been isolated.  Hence, it was immediately clear that
three generations were needed for baryogenesis.

And so, from our present perspective, if the number of generations
is a random field that ``freezes out'' in the early universe,
the ``selection requirement'' that our universe contain physicists
strictly imposes $N_{\rm gen}>2$.

\section{{A Testable Anthropic Prediction}
\label{sec:prediction}}

Why then are there only three generations?  Within the anthropic-principle
framework, the answer is clear:  the random ``generation number'' field
has a steeply falling probability of freezing out with increasing value
of $N_{\rm gen}$.  Hence, my prediction is that when the theory of
these fields is developed, it will be found that the probability
of high $N_{\rm gen}$ is small, perhaps because extra generations
require the mediation of a high-mass (therefore heavily suppressed)
particle.

Although Sakharov's 3 conditions for baryogenesis were inspired by
the discovery of CP violation in the quark sector, the actual channel
for baryogenesis is not yet established and therefore may involve
other particle sectors.  For example, one
possibility is that the earliest particle asymmetry 
is leptogenesis through the neutrino sector (rather than
baryogenesis directly through the quark sector), and this 
indirectly induces baryogenesis by processes that conserve
$B-L$ (baryon minus lepton number) but violate each separately,
by converting anti-leptons into baryons.  If neutrinos are Majorana
particles, and so their own anti-particles, then it would be
possible to have CP violation with only two generations.  In this
case, the appearance of a third generation would be superfluous
from the standpoint of
human existence, so that no such anthropic argument could be
made.  This serves to underline that anthropic arguments in general
must be based on a thorough understanding of the physics of our
universe.

\section{{A Quantitative Example}
\label{sec:example}}

Let us suppose that, some time in the future, it is firmly established 
that baryogenesis is due to quark-sector CP violation.  And further,
that continuing searches for heavy quarks and leptons at LHC fail
to find a fourth generation, thus tending to confirm the present 
conclusion that $N_{\rm gen}=3$.  And finally, that physicists converge
on a theory of generation-number ``freeze out'' with probability
$P\propto N_{\rm gen}^{-\alpha}$, where $\alpha$ is established
to be some definite number.

If $\alpha=1.05$, then the probability of our universe having
exactly $N_{\rm gen}=3$ (versus $N_{\rm gen}>3$) would be
$\sim 1/60$.  This would not, by itself, rule out the multiverse,
because events of this level of improbability do happen.  But
it would by itself be reason for extreme caution.  And if the
multiverse failed a few such tests, it would be ruled out.

On the other hand, if $\alpha=10$, then the prior probability
of the observed $N_{\rm gen}=3$ would be 94\%, which would
be consistent with the multiverse.  Of course, scientific hypotheses
can never be finally ``proved'', but if the multiverse passed
many such tests, it would come to be accepted by the same
process as other theories.

\acknowledgments

I thank John Beacom and David Weinberg for useful discussions,
and Basudeb Dasgupta for pointing out the possibility of leptogenesis
path to baryogenesis.


\end{document}